\def\year{2020}\relax

\documentclass[sigconf]{acmart}

\usepackage{multicol}
\usepackage[utf8]{inputenc}
\usepackage{blindtext}
\usepackage[T1]{fontenc}
\usepackage{natbib}
\usepackage{graphics}
\usepackage{hyperref}
\usepackage{dcolumn}
\newcolumntype{d}[1]{D{.}{.}{-1}}
\newcolumntype{n}[1]{D{,}{,}{-1}}
\usepackage{pdflscape}
\usepackage{balance}
\definecolor{mylinkcolor}{RGB}{0,0,0}
\hypersetup{colorlinks,allcolors=mylinkcolor,citecolor=mylinkcolor}
\usepackage[capitalize]{cleveref}

\allowdisplaybreaks
\newcommand{\xhdr}[1]{\paragraph*{\bf #1}}

\title{The dynamics of U.S. college network formation on Facebook}

\author{Jan Overgoor}
  \affiliation{\institution{Facebook}}
  \email{janovergoor@gmail.com}
\author{Bogdan State}
  \affiliation{\institution{Facebook}}
  \email{bogdanstate@gmail.com}
\author{Lada A. Adamic}
  \affiliation{\institution{Facebook}}
  \email{ladamic@fb.com}

\begin{document}

\begin{abstract}
In the U.S., a significant portion of many people's life-long social networks is formed in college. 
Yet our understanding of many aspects of this formation process,
such as the role of time variation, heterogeneity between educational contexts, and the persistence of ties formed during college,
is incomplete.
In order to help fill some of these gaps, we use a population-level dataset of the social networks of 1,181 U.S. institutions of higher education,
  ranging from 2008 to 2019, to provide a detailed view of how the structure of college networks changes over time.
The most prominent feature in the evolution of these networks is the burst in friending activity when students first enter college. Ties formed during this period play a strong role in shaping the 
structure of the networks overall and the students' position within them. Subsequent starts and breaks from instruction further affect the volume of new tie formation. 
Homophily in tie formation likewise shows variation in time.
Same-gender ties are more likely to form when students settle into housing,
while sharing a major spurs friendships as students progress through their degree. 
Properties of the college, such as whether many students live on campus, also modulate these effects.
Ties that form in different contexts and at different points in students' college lives vary in their
  likelihood of remaining close years after graduation. Together, these findings suggest that educational context mediates network formation in multiple different ways.
\end{abstract}

\maketitle

\section{Introduction}

Social networks represent a fundamental part of the structure of societies. Many such networks are created in organizational contexts, and understanding their formation and evolution dynamics is key to understanding social change itself. Given their strong effects on shaping the life-course, ties formed during college are particularly important to study.
More than half of the U.S. adult population has attended a higher education institution \cite{ryan16}. Colleges are thought of as engines of social mobility \cite{chetty17}, but they also contribute to processes that play important roles in the reproduction of economic inequality. It is common, for instance, for people to find their partners or close friends during college \cite{arum08}, which contributes to social stratification in U.S. society more generally \cite{gerber08}. In addition to improving human capital, university studies also have important effects on social capital. For many, college is the first exposure to a larger and more diverse social environment than where they grew up, which supports later access to the labor market \cite{granovetter73}.

\begin{figure}
  \centering
  \includegraphics[width=\columnwidth]{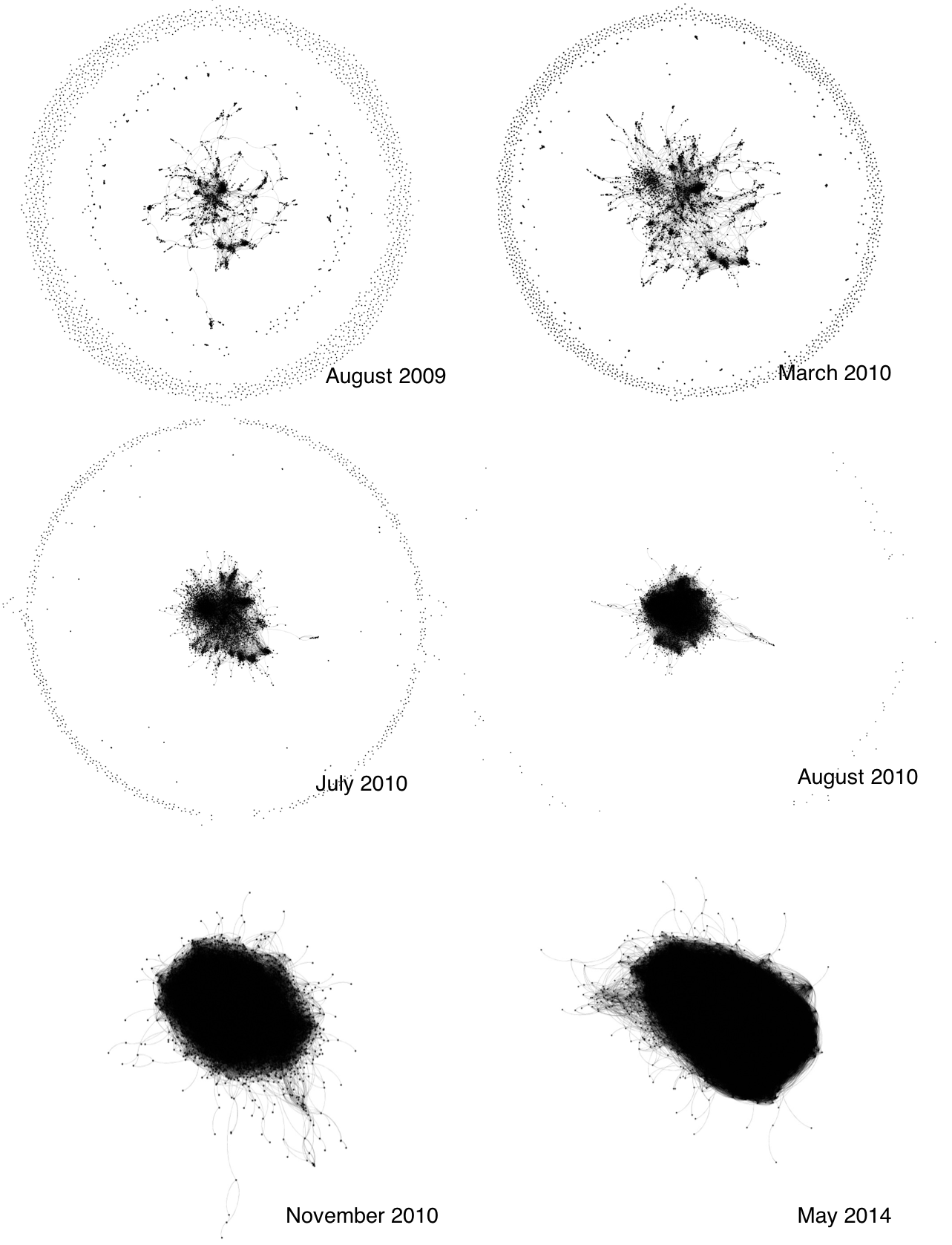}
  \caption{
      The evolution of the network of Facebook friendships between students entering a state college in 2010. Prior to August of 2010, small clusters likely represent friends from the same home town and high school. As the very first semester starts, an avalanche of new friendships forms. 
     }
  \label{fig:gephi_over_time}
\end{figure}

Higher education carries capital importance to both individual and societal outcomes, for which social networks are an important mediator. Perhaps because of this, and perhaps because of the availability of undergraduate students as a study population to social scientists, many studies have been conducted about the structure and formation of college social networks, which have been reviewed in \cite{biancani13}. 
Collecting social network data, even within a single class, is a daunting task \cite{mcfarland01}.
This has led to most prior studies being focused on one or a few cohorts at one or a few schools, and often using at most few static snapshots of data. While surveying students provides a rich set of insights within a particular college, there has been little opportunity to study variation between colleges.
Data from online social networks helps address some of these limitations~\cite{traud11}, especially those concerning scale and temporal dynamics. However even with online data, at most a handful of schools have been studied at a time.
Because Facebook has historically been popular with college students (the platform having originally only been available on US college campuses),
  it provides a natural context to study the online social networks of college students.
Figure~\ref{fig:gephi_over_time} visualizes the evolution 
  of the network of Facebook friendship ties between people who entered a state college as first-year students in 2010. Although some students were acquainted prior to enrolling,
  that small number of ties is quickly eclipsed by the many new ties that are created within the first weeks of college. It is the dynamic formation of these 
  ties that is the subject of this paper. 
   
Using a population-level dataset of the social networks of 7,586 distinct class cohorts associated with 1,181 U.S. higher education institutions \cite{overgoor19b},
  we aim to address three gaps in the current literature.
First, knowing the timing of the creation of social ties allows us to go beyond inference based on snapshots of data.
We describe how the structure of college networks change over time, from the start of classes throughout the expected graduation date.
Second, we address heterogeneity of network structure by college characteristics.
Different schools provide different ecologies for students to form their networks in \cite{mcfarland14,stevens08}.
This can come from the demographic composition of the student body,
  or the specific logistical and physical factors of the institution.
Finally, we look at the persistence of social ties formed in college,
  by employing a measure of tie closeness in the present day.

We find that tie formation patterns, both in volume and in kind, vary through time. Changes in
  how social ties form correspond to major events throughout the college tenure. 
  The network changes the most during the start of school,
  the start of a new academic period, and during the ``rush'' period for fraternities and sororities.
These trends also vary by school type.
For example, students attending school far from home make more same-year and fewer same-home town connections
  when school starts. School type also mediates the role played by shared characteristics in the formation of new ties.
Homophily by gender is lower in Historically Black Colleges and Universities (HBCUs) and higher in schools with high participation in Greek organizations,
  especially during recruitment times.
These observations imply that both the timing and context of a network sample have a potentially large effect on what data is gathered.

While the characteristics of new ties formed vary depending on when they are formed, the rush of friending activity at the start of college leads to
the overall structure of each cohort-level social network developing and stabilizing relatively quickly. This is 
reflected in the size of the giant component, the average shortest path, clustering coefficient and maximal modularity.
Similarly, since many ties are formed around the time of starting college, an
individual's overall position in the network -- as measured by eigenvector centrality -- typically does not change much after the start of school.

Structural stability should not be taken to minimize the importance of the time aspect for the development of individual ties.
Our results also reveal that both the time and place in which a tie was formed affect how long the tie stays relevant after school.
Same gender ties and ties formed in commuter schools are relatively more likely to last.
Our results provide clear evidence that educational context mediates the formation of social networks in college, and offers
a glimpse at the typical rhythms in which the social fabric that ties together a cohort of individuals is established and develops.

\subsection{Related Work}

Interest in the evolution of social networks dates back to the earliest days of social science,
  and much of the early work was about schools \cite{hanifan16,french48,moreno34}.
More recently, data for studies of network formation during college have generally come from one of three sources of data:
  in-person surveys \cite{perl88,vanduijn03,godley08,stadtfeld18},
  email data \cite{marmaros06,kossinets06,kossinets09},
  and friendship data from Facebook \cite{lewis08,wimmer10,traud11,lewis12,jacobs15,lewis18}.
All three data sources have their own affordances and limitations.
Survey data can be tailored to the specific research question, but is expensive to gather and suffers from response bias.
It is generally not possible to observe events as they happen,
  so instead inference has to be based on one or multiple snapshots of the network \cite{snijders01,snijders06}.
Digital trace data often does contain individual events and their timing,
  and can provide a complete record of data within the specific digital platform.
However, digital data is often a convenience sample,
  and research is thus limited by what data is recorded and confounded by usage of the platform \cite{lewis08}.
This applies to our study as well, but is mediated by the high adoption of the Facebook platform within the population of U.S. college students.

Two common recurring themes in network formation during college are propinquity and homophily.
Propinquity is the tendency for social networks to be spatially organized,
  with proximity a key factor influencing the likelihood of social tie formation.
This is relevant with respect to shared foci,
  like dormitories \citep{festinger50,marmaros06,lewis12,vanduijn03},
  classes \citep{kossinets06},
  and extracurricular activities \citep{vanduijn03,schaefer11}.
Homophily is the preference of associating with others who are similar \citep{mcpherson01,currarini10},
  and is particularly prominent in educational settings,
  where it occurs along dimensions like race \citep{marmaros06,godley08,wimmer10},
  gender \cite{vanduijn03},
  and socio-economic status \cite{lewis12}.
Other related factors include personality and the social climate of one's residence \cite{perl88}. 
Structurally, triadic closure (befriending friends-of-friends) is a common factor in formation \cite{vanduijn03,jackson07,wimmer10,lewis18}.
A more long-term perspective found that ties formed in school can have an effect of one's social network even 20 years later \cite{burt01}.

Due to the sparse availability of data that is both granular and longitudinal,
  there has been little work on whether and how formation dynamics \textit{change} over time.
Early work argued that physical proximity and visible similarity matter early on in college,
  while later network structure matters more \cite{vanduijn03}.
The decline in the relative importance of homophily was also documented in more recent work \cite{godley08,lewis18}.
Another study found no change over time \cite{stadtfeld18}.

Most work on social networks in college has focused on single institutions, with some exceptions \cite{traud11,overgoor19b}.
In contrast, the comparative study of social networks in high schools has advanced more,
  because of the longitudinal ``ADD Health'' dataset \citep{harris08},
  which collected data from over 90,000 individuals who were enrolled in middle school or high school in the US during the 1994-95 academic year.
This includes research on homophily \citep{joyner00},
  structural dynamics \cite{goodreau09},
  propinquity of extra-curricular activities \citep{schaefer11}, 
  and the relation of various behaviors to network structure.
One paper takes a similar approach to ours, by relating structural elements of high school networks to school covariates \cite{mcfarland14}.

\section{Data}

The college networks dataset was described in earlier work \cite{overgoor19b}, but we briefly review it here. The data covers friendships between currently active users of the Facebook platform in the United States, who reported college attendance on their profile.  All data was de-identified and analyzed in aggregate. 
It combines
  self-reports of demographic information on Facebook,
  information about Facebook friendships,
  and institution-level data about colleges in the United States.
  
  Users of the Facebook platform can self-report demographic information about school attendance, age, and hometown.
Self-reports of higher education attendance are accepted,
  if they can be resolved against a known U.S. higher education institution,
  and if a sufficient number ($n=10$) of one's Facebook friends
  have likewise been identified as attending the same higher education institution.
  Ties between students are represented by Facebook friendships,
  which are initiated by one of the two people by sending a ``friend request'' through the platform,
  and accepted by the other party.
  
Institutions of higher education in the United States are enumerated in the College Scorecard dataset released by the U.S. Department of Education.\footnote{Data can be downloaded from the College Scorecard website, \url{https://ed-public-download.app.cloud.gov/downloads/CollegeScorecard_Raw_Data.zip}}
This dataset provides school characteristics like class size, whether the school is a public or private institution,
  the admission and graduation rates, and whether the school predominantly serves a minority population.
Other derived school characteristics include whether students primarily commute to the school, and the share of students that participate in Greek life.
People who reported having attended a school, but did not report a starting year,
  are assigned to an entry-year class
  using a classification algorithm trained on the data of people who did report a starting year.
Only predictions with a sufficiently high confidence (75\%) are used.
Only entry-year classes with a size close to the reported class size (within a factor of 0.5 to 1.25) are kept for analysis.

The dataset contains 224.3 million within-cohort friendship ties between 6.9 million users assigned to 7,586 entry year cohorts in 1,181 U.S. institutions of higher education. A further 288.3 million edges occur at the same institution, but between cohorts. Of the 1,181 institutions in the sample, according to the Carnegie Classification,\footnote{\citep{carnegie15}, retrieved from \url{http://carnegieclassifications.iu.edu}} 68 are Historically Black colleges and universities, 31 are women's-only institutions, 65 are classified as Hispanic-Serving Institutions, and 227 are undergraduate-only institutions.
On average, 19.8\% of one's Facebook friendships are with others who went to the same school.

Although the data set is large and has broad coverage of colleges, it may be biased in several ways. First, it is limited to people who use Facebook. Despite this, we find that the resulting classes are similar in size and demographics (gender, age, and within-state) to those as reported in external data.\footnote{Data as reported in the Enrollment and Employees in Post-secondary Institutions report as released yearly by the NCES.} Second, the creation of a friendship on Facebook does not always exactly correspond with two people becoming friends offline.
People can add each other long after (or before) they have actually met,
  and the threshold at which people consider themselves friends on Facebook varies per individual \cite{zywica08}.
However, in the restricted context of college
  the interpretation of a Facebook friendship as signifying an in-person meeting becomes more viable.
Facebook usage among sampled college students was upwards of 95\% \cite{ellison07,lewis08},
  and college students report in surveys that only a very small percent of their Facebook connections are online-only \cite{ellison07,mayer08}. 

\subsection{The structure of college networks}
The 1,181 school graphs vary in their structure according to characteristics of a college, such as the size of its student body.
Typically people in larger school graphs have a higher average degree, up to an average of about 150 friends, corresponding to other observations about cognitive limits to social network size \cite{hill03,gonccalves11}. It is worth noting in the context of node degree having an effective ceiling, edge density will decrease with graph size,
given that this statistic relies on a denominator proportional to the square of the number of nodes in the graph.
Similarly, the average local clustering coefficient also decreases in larger networks, as your friends are less likely to be friends \citep{leskovec08}.
Larger graphs also have a higher modularity.

\section{Trends in tie formation}
We first describe a number of trends in the structure of the college graphs over time.
In each case, we map events relative to when the individual started college.
We look at the six years starting one year before college starts to one year after it ends.

\begin{figure}
  \centering
  \includegraphics[width=\columnwidth]{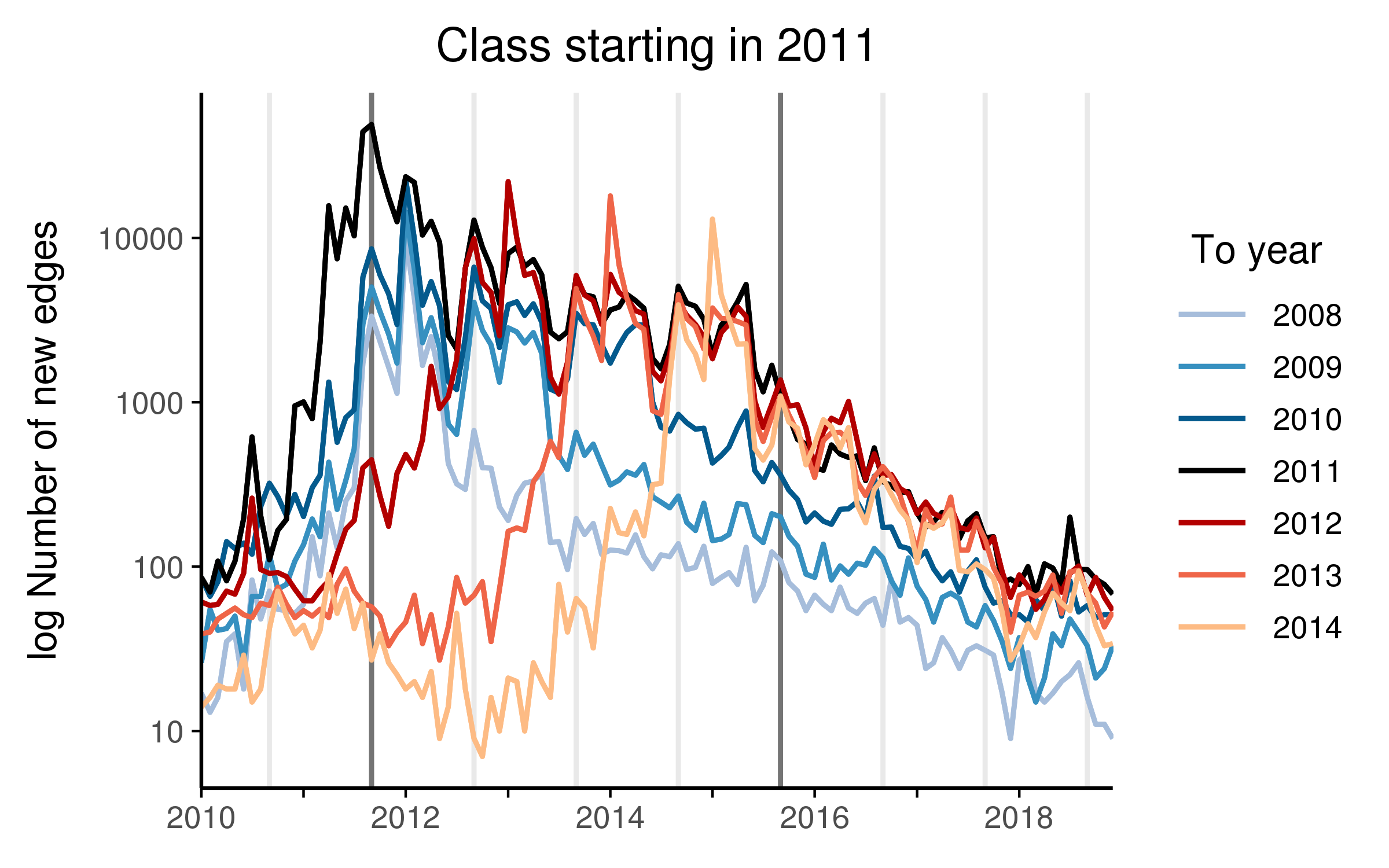}
  \caption{
    Monthly friending volume for a cohort of students of a selective private university who enrolled in 2011.
    The data are grouped by the entry-year class of the friend.
    Vertical lines mark the beginning of each new \textit{school} year.
  }
  \label{fig:crossyear}
\end{figure}

To gain intuition, we'll first look at intra- and inter-cohort friending for students entering a selective private university in the northeast in 2011, illustrated in Figure \ref{fig:crossyear}.
Friending starts ramping up early in 2011, likely after students have received their admission decision. Once enrolled, a flurry of friending activity occurs. Freshmen predominantly befriend those in the same class, but also older cohorts who are already at the college. 
With the start of every subsequent school year, as a new cohort enrolls, friending with the new cohort outpaces friending within the 2011 cohort. At the same time, the rate of new tie formation falls between the 2011 cohort and the cohort that has graduated.
After four years, with many of the students having left the college, befriending tapers off.

\begin{figure}
  \centering
  \includegraphics[width=\columnwidth]{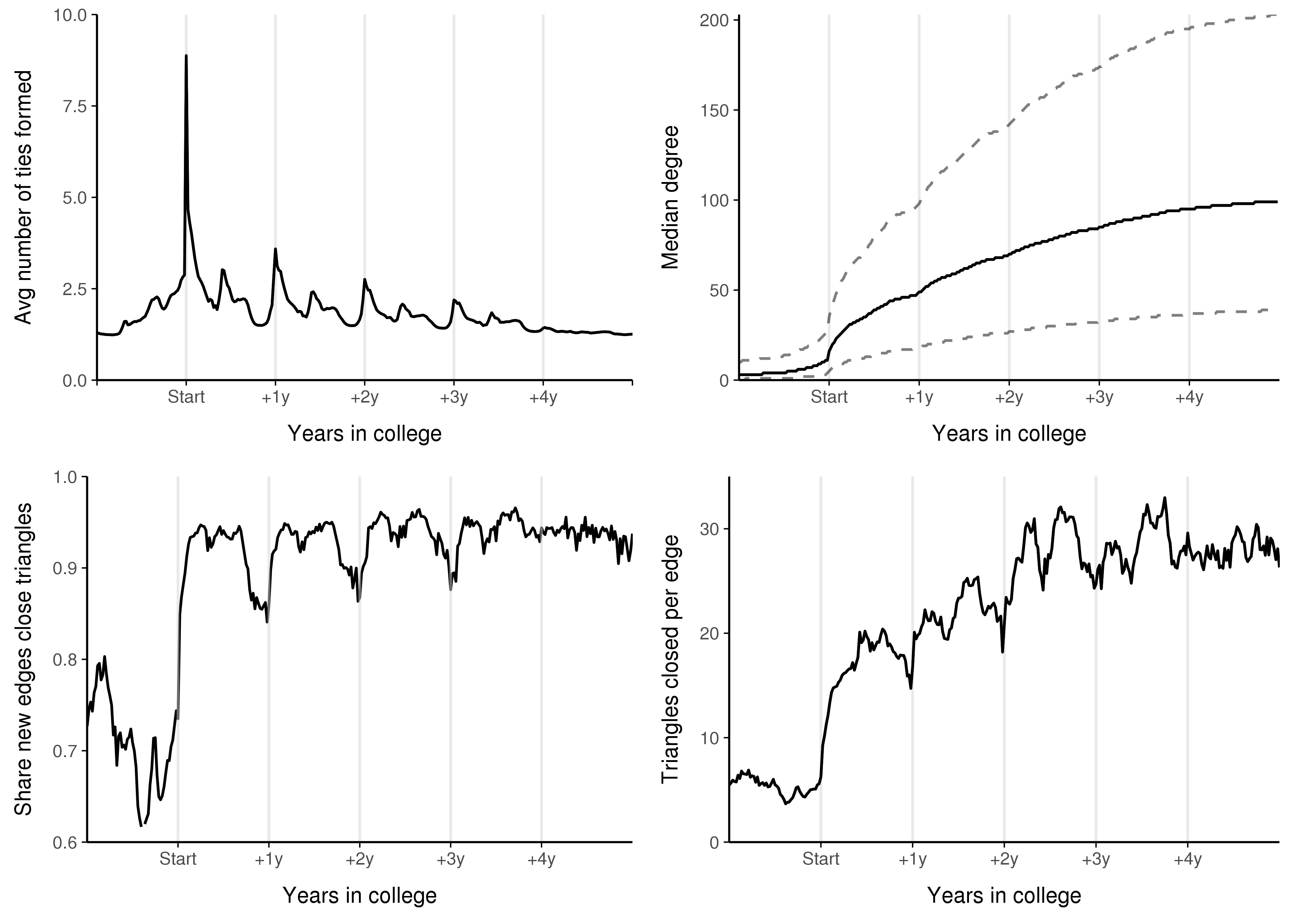}
  \caption{
    Characteristics of network growth across all schools and entry-year classes, plotted by week relative to initial enrollment (start).
    Top left: average number of edges formed per week.
    Top right: median degree. Also shown are the 25th and 75th percentiles.
    Bottom left: share of new edges that close triangles per week.
    Bottom right: average number of triangles closed by new edges per week.
  }
  \label{fig:new_edges}
\end{figure}

Next we'll consider tie formation patterns across all colleges and all cohorts.
\xhdr{Volume} We start with the timing of when during college Facebook friendships tend to form.
In Figure \ref{fig:new_edges} (top left panel), we plot the average number of new within-college ties created each week.
The number of new ties starts ramping up slowly leading up to the beginning of school,
  and spikes in the first week of school.
As the years goes on, the average number of new ties decreases steadily,
  with pronounced jumps at the beginning of each school year,
  and to a lesser extent mid-school year (in January).
These moments correspond to when classes start,
  and for those in schools that offer dormitories, moving into a new residency.
After four years, the volume of within-college ties decreases and stops showing the seasonal effect.

The top right panel shows changes in the median degree, showing the accumulation of new ties over time.
The median degree at the start of school, considering all people in our sample, is 10.
Within two months it increases to 27, at the end of the first school year it is about 50, and by the end of school it is 95.
We also plot the 25th and 75th percentiles of degree over time, which shows that new edges are not evenly distributed, even later in school.
The 25th percentile grows only slowly, while the 75th percentile grows much faster to about 200 at the end of school.

\xhdr{Triadic closure} We know that triadic closure -- the creation of ties between individuals who share friends in common -- is an important driver in making social connections.
Friends of friends are pre-vetted,
  and one is more likely to meet them in shared social and physical spaces \cite{festinger50,vanduijn03,jackson07}.
Given that the social networks in the college networks data set are so dense (the average edge density is 0.03),
  most new ties are necessarily going to close triangles.
However, as shown in Figure~\ref{fig:new_edges} (bottom left panel), there is a strongly seasonal component to this trend as well.
We plot the share of new edges that have at least one friend in common (i.e. close at least one triangle) \textit{within the same school}, by week.
Edges forming in the half year before the start of school are less likely to be between friends of friends than those forming prior to that six month period.
This is a time when social networks are in a state of flux, when a lot of new introductions occur between future friends and acquaintances. 
As school starts, the rate of triadic closure quickly reaches its steady point of about 0.95.
Then, every year during the summer, there is a drop in ``friend-of-friending,''
  and so people are more likely to befriend those outside of their immediate social circles.
  
We hypothesize that this drop-off in triadic closure during the summer is due to the fact the social context of college changes when school is not in session.
This is the period when students travel, go back home, or engage in summer employment on- or off-campus. Any ties that are created at this point may be more
likely to result from contexts different to the school-year one: for instance, students who happen to work on-campus summer jobs may all be housed in the
same few summer dormitories at the college, with ties formed during the summer thus being more likely to be bridging. Further evidence in support of this explanation
can be observed in the smaller decline in triadic closure that occurs around 3-4 months into the school year, a period coinciding with Winter break, a period
which likewise may produce less socially-embedded friendships and acquaintances.

Interestingly, triadic closure remains high even after the estimated end of the time at school -- this is likely a result of friendship circles established during college ``filling in,''
either as new users join Facebook after college, or when pre-existing friends find one another on Facebook, or through social situations facilitating the introduction of
those classmates who, despite having many ties in common, had never actually met during their college days.

A similar pattern is shown in the bottom right panel of Figure~\ref{fig:new_edges}, where we plot the average \textit{number} of triangles that are closed with every new edge. As expected, this trend is mostly increasing and has a logarithmic shape. Especially later in school tenure, there is a pronounced pattern of heightened friend of friending at the end of each academic period.

\xhdr{Homophily}
Homophily -- the positive association between shared social characteristics and likelihood of tie formation --  is frequently an important factor in social networks \cite{mcpherson01},
  and especially so for social connections formed in college \cite{vanduijn03,godley08,wimmer10,lewis12}.
In prior analysis of this data, homophily by year was found to be biggest in more selective schools and HBCUs \cite{overgoor19b}.
Gender homophily was highest in religious schools and in schools with more Greek life participation, and lowest in HBCUs.
Here we look at the role homophily plays in edge formation over time.
In order to account for availability and the changing composition of the student body with each incoming and outgoing class,
  we use a modified version of Newman's Homophily coefficient \cite{newman03}:

\[ H^t = \frac{\sum_i e^t_{ii} - \sum_i a^t_i b^t_i}{1 - \sum_i a^t_i b^t_i } \]

In this modified version of the original formulation, the term $e^t_{ii}$ for a specific class and time slice $t$,
  is the share of edges during that time and involving at least one member of that class,
  that share a particular feature $i$.
The term $a^t_i$ refers to the share of all edges during time $t$, where the member of the class of focus has feature $i$.
In contrast, $b^t_i$ is the share of all edges during time $t$, where either node has feature $i$.
This formulation retains the interpretation of the homophily coefficient as the observed same-feature edges,
  as compared to the expected same-feature edges, if there was no same-feature bias. 
We look at homophily across four dimensions: gender, entry year class, major, and home town.

Overall, gender homophily is positive throughout college, especially immediately prior to enrollment and during the friending burst at enrollment. It subsequently drops, but sees recurring spikes twice each year. 
As will be shown in the next section, this is partially due to the start of the the academic year,
  but mostly driven by the rush periods of Greek organizations.
Homophily by year is also at its highest point right at the beginning of school.
This makes sense, as students are most likely to meet others from the same class due to shared social contexts like
  introduction weeks or, in the case of residential colleges, freshmen dormitories.
Every subsequent year, the average homophily by year drops, likely linked to friending with new cohorts, as we saw in Figure~\ref{fig:crossyear}.
In contrast, homophily by major increases steadily during the progression of college, with the exception of summers, when students may not be in class.
The most easily explained trend is in homophily by home town. Prior to the start of college, homophily by home town is the strongest absolute factor, but this drops immediately at the start of college. Very small peaks repeat each summer, as students at residential colleges may temporarily return to their hometowns. After school ends, the trend starts slightly increasing again.

\begin{figure}
  \centering
  \includegraphics[width=\columnwidth]{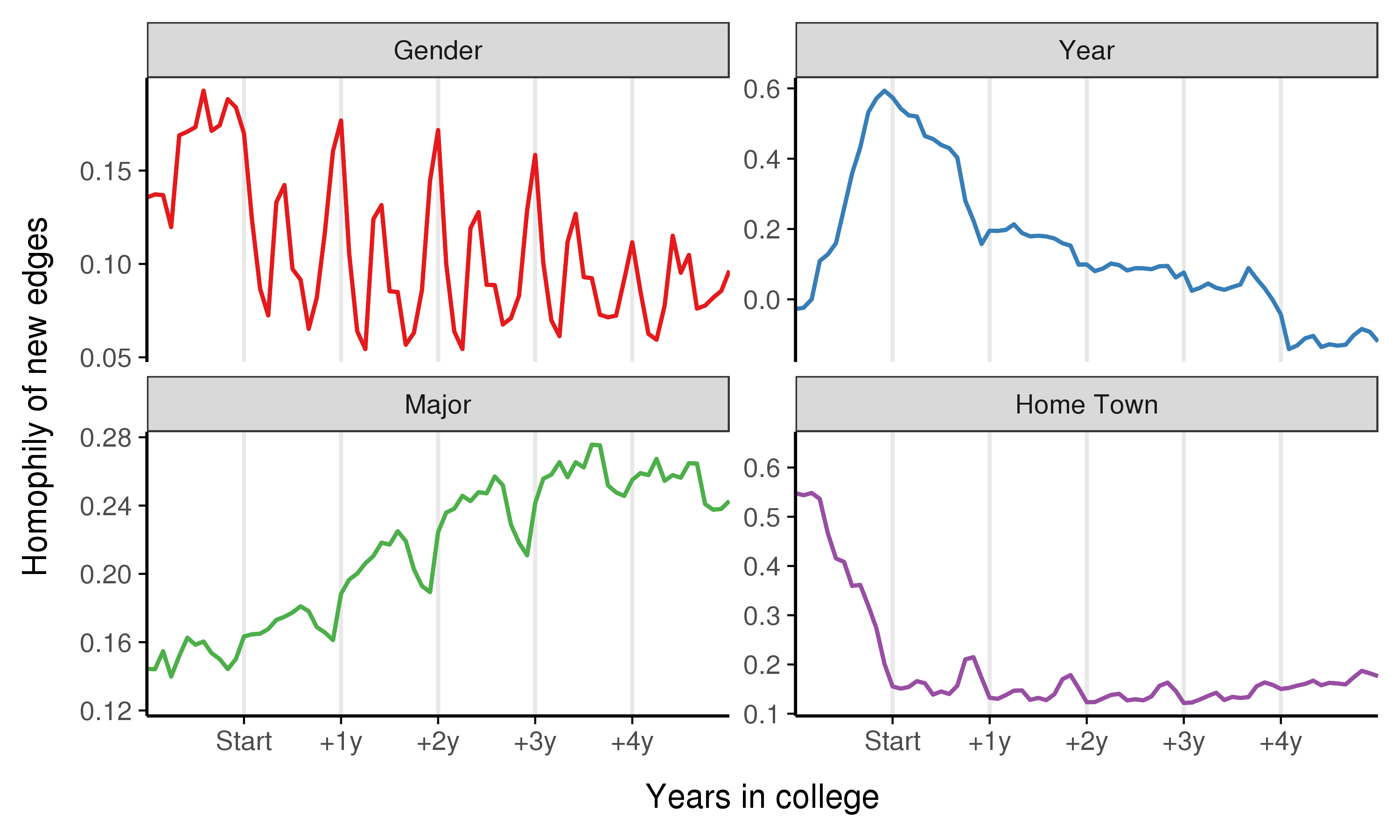}
  \caption{
    For each of four demographic dimensions, the homophily coefficient $H^t$ of new ties formed in month $t$.
    The data are computed per month since starting college, and averaged over all college classes.
    The y-axes have different ranges. 
  }
  \label{fig:homo_avg}
\end{figure}

\begin{figure}
  \centering
  \includegraphics[width=\columnwidth]{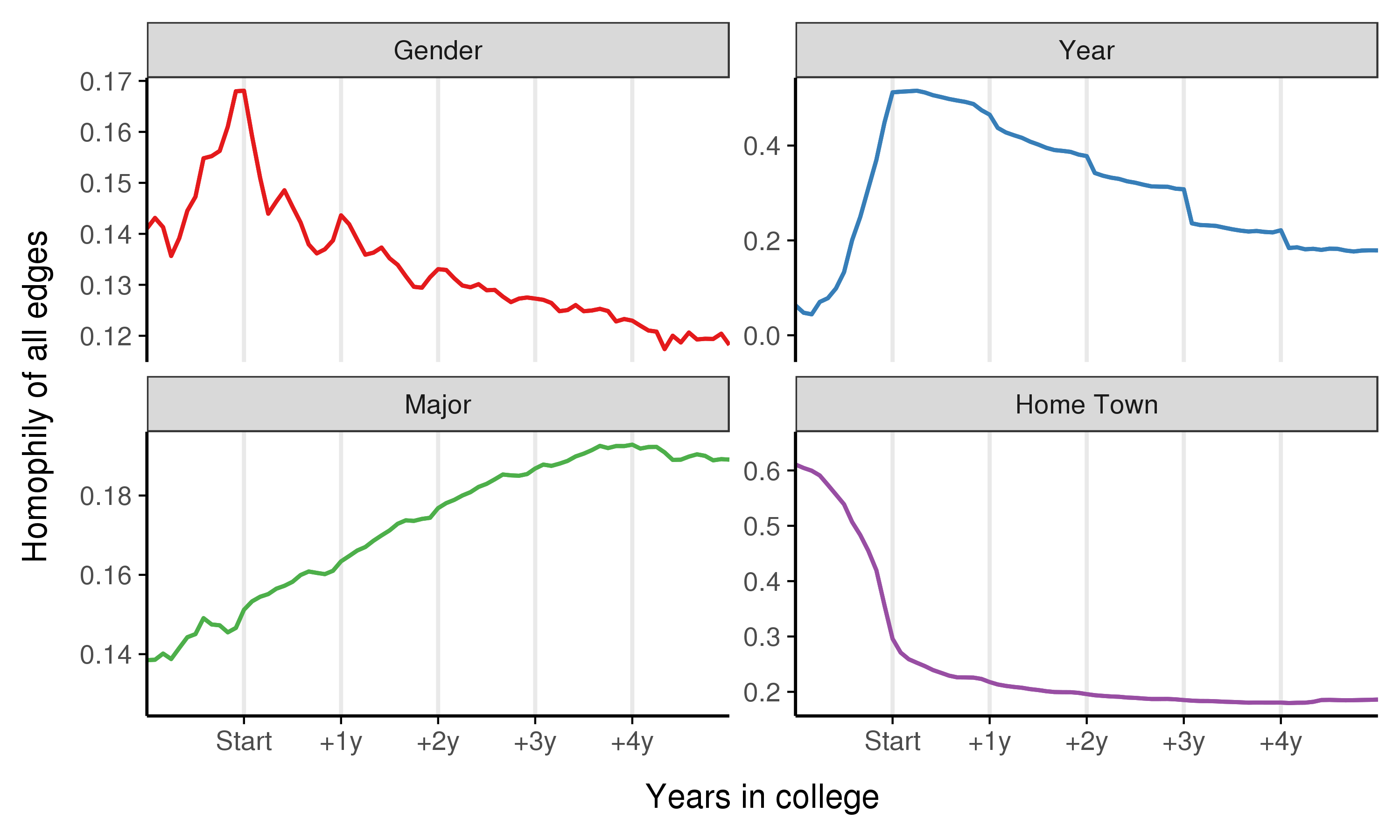}
  \caption{
    For each of four demographic dimensions, the homophily coefficient $H^t$ of all ties formed until month $t$.
    The data are computed per month since starting college, and averaged over all college classes.
  }
  \label{fig:homo_avg_cum}
\end{figure}

In Figure \ref{fig:homo_avg_cum} we show the cumulative measure of homophily in the network up to time $t$. It highlights that overall homophily in the network is strongly influenced by the peak in friending activity at the start of the first year, with drift according to subsequent increases or decreases in homophily.

\xhdr{Homophily by school type}
Next, we look at how these trends in homophily of new edges vary by school type.
To do so, we fit a separate regression model for each of the four dimensions.
Each model has the already discussed measure of homophily of new edges per month
  as the dependent variable,
  and every month (since starting college) interacted with various covariates
  about the college as independent variables.
This model wrongly assumes that the relative ``effect'' of school type on homophily is
  independent for each month, which is not the case.
However, we use this model to highlight seasonal differences in edge forming behavior
  as averaged across school types.
Each data point is thus the homophily per dimension for a specific class in a specific month since starting college.

\begin{figure}
  \centering
  \includegraphics[width=\columnwidth]{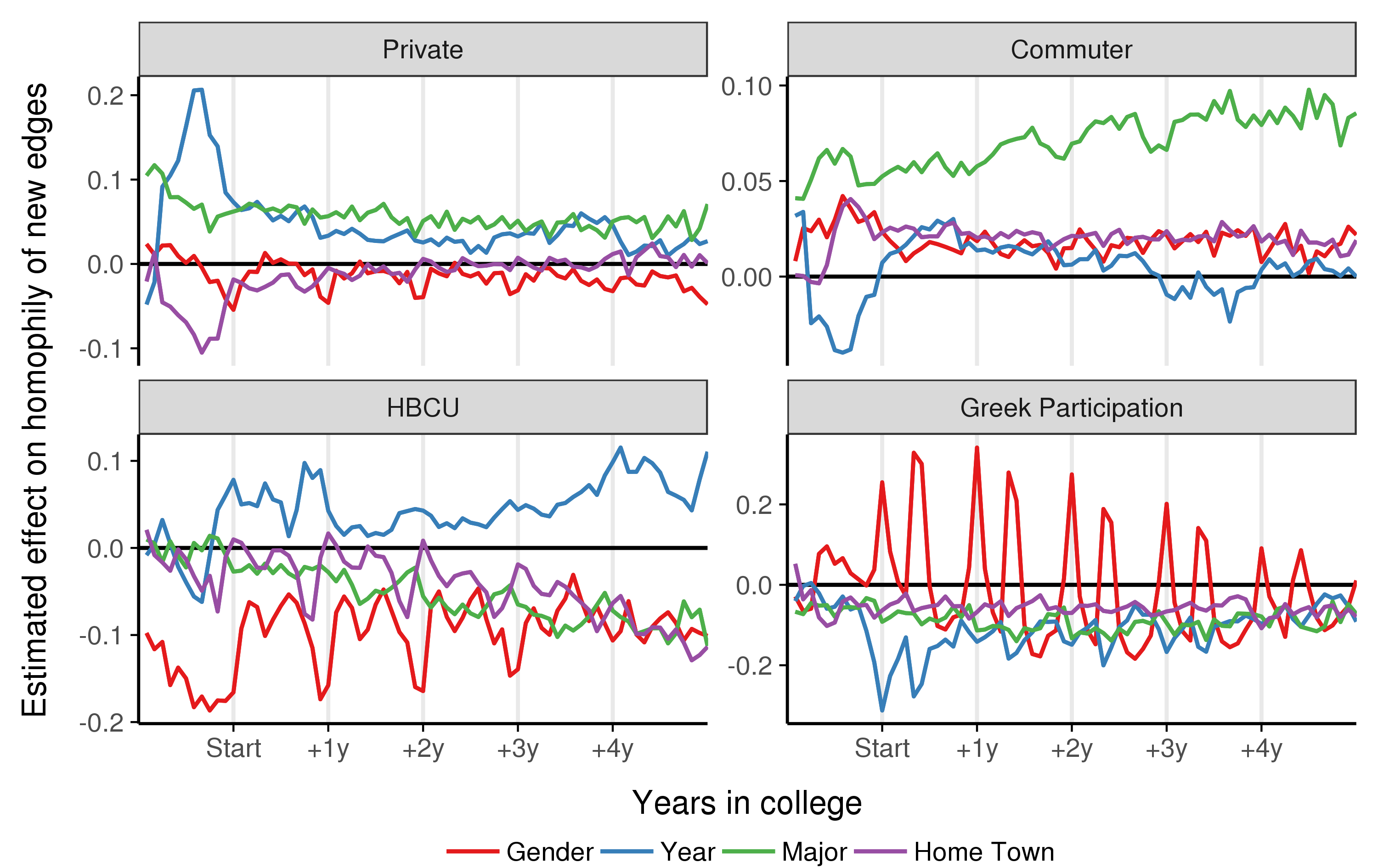}
  \caption{
    Estimated difference in homophily of new edges, by various school types and for one of four different dimensions.
    The plotted estimates are the results of a regression model,
      with covariates for other (not shown) school features including class size, graduation rate, whether the school is religious, or serves a single gender.
  }
  \label{fig:homo_all}
\end{figure}

The resulting estimates are plotted in Figure \ref{fig:homo_all}. We go through them one by one.
New edges in private schools show consistently more homophily with respect to year and major, with a big spike in the period before school starts.
New ties formed pre-college are approximately 20\% more within the same-year for private schools, as compared with public schools.
This outcome may be the result of private schools having more extensive admission-related events that also result in the formation of social ties between future students.
Similarly, pre-school edges are significantly \textit{less} likely to be from the same home town for private schools.
Commuter schools show the opposite effect, with pre-college edges being less likely to be same-year and more likely same-hometown.
  In general, commuter schools have a small but significant positive effect for home town homophily.
Most commuter schools are public schools,
  but since both covariates are included in the same model,
  these effects display the marginal effect of each feature separately.

With respect to gender homophily, Historically Black colleges and universities (HBCUs) and schools with a large Greek participation have more distinctive patterns.
HBCUs have strongly more gender mixing (ranging between 10 to 20\%),
  especially during the beginning of the school year.
The spiky pattern in gender homophily we saw in Figure \ref{fig:homo_avg}
  appears to come mostly from schools with high Greek participation.
The covariate for Greek participation is a rate, so the plotted effect sizes represent the effect for schools with a 100\% Greek participation rate.\footnote{In our data, schools with a high rate have a rate of about 50\%, so the effect size is about half of what is shown, and lower for schools with a lower rate.}
New edges in these ``Greek'' schools are strongly same-gender during the beginning
  of the school year, and in the beginning of the calendar year. These times correspond to when most schools have their rush (initiation to Greek organizations) time. Some schools have a single rush period, others have two. Since Greek organizations are mostly not co-ed, students predominantly meet others from the same gender during this time.
Outside of rush time,
  edge formation in schools with a strong Greek presence exhibits \textit{less} homophily by gender.
However, overall (cumulative) homophily by gender is still higher in Greek schools.
Year-homophily is also lower in schools with high Greek participation,
  especially in the first year and during rush times.

\section{Trends in structure}
\begin{figure}
  \centering
  \includegraphics[width=\columnwidth]{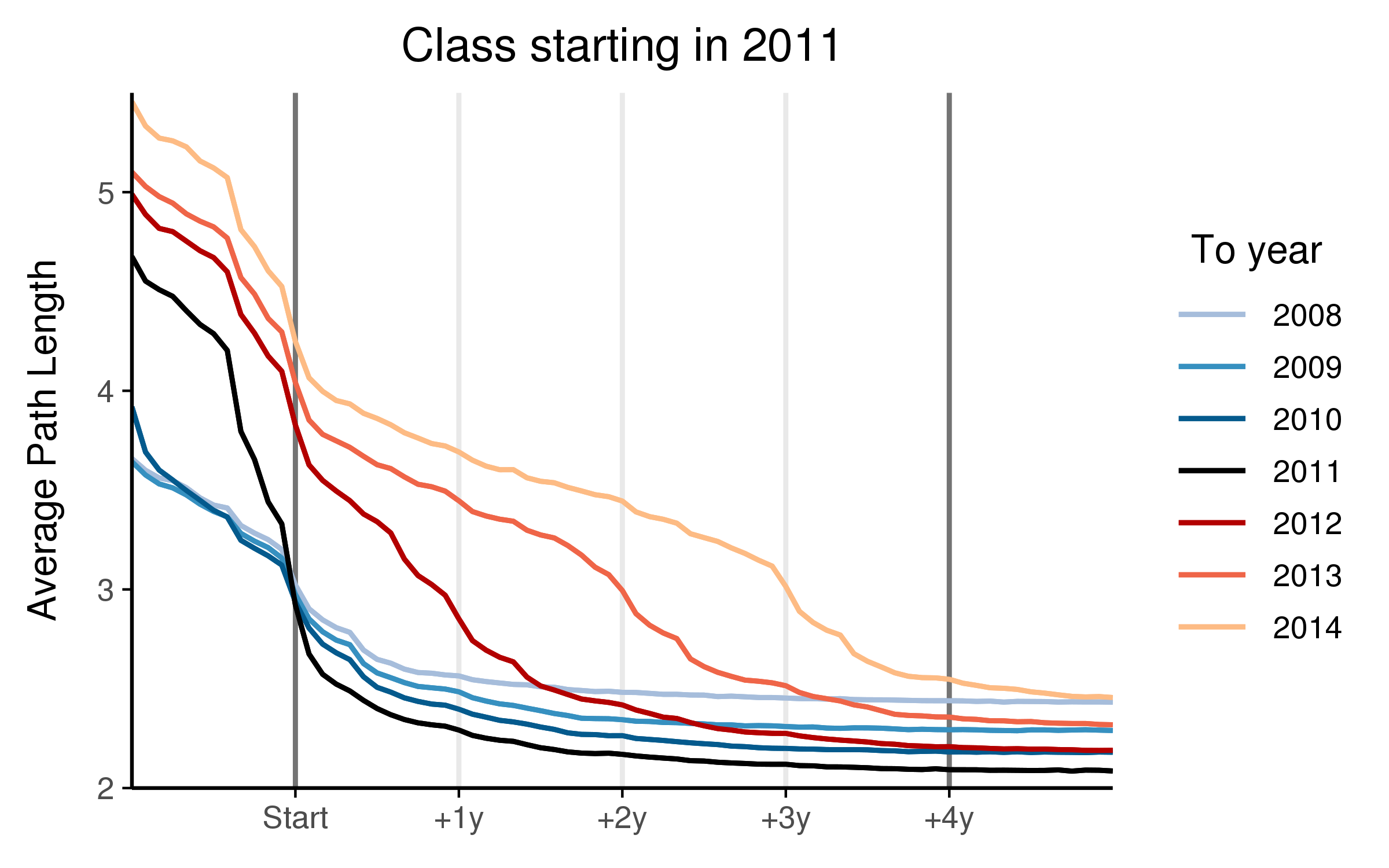}
  \caption{
    The average shortest path length between students of a private university starting in 2011, and students starting in other years.
    The data are computed for every month.
  }
  \label{fig:path}
\end{figure}

Next we move from observing pairwise tie formation to describing the evolution of the entire network structure.
As an example, consider the average shortest path length in the school graph of a selective private university over time between students starting in 2011 and students from other classes. From Figure \ref{fig:path},
a number of intuitive observations can be made. Before physically joining the school, incoming students on average are more proximately connected to those who are already in school than they are in to their own cohort. This may be due to students who are already at the college being part of a dense network, so connecting to any one current student provides indirect connection to anyone else in the school. Right before school starts, the average shortest path length drops significantly as many new edges get made. Students are most closely connected to others from the same entry-year class and stay so throughout their tenure. As new classes come in, their network distance drops in a similar fashion. However, the long-run average distance depends on the time difference. People that started in 2011 are, 5 years after starting college, on average as close to those who started in 2010 as those who started in 2012.

Given the rapid convergence of the average shortest path within a cohort, we are particularly interested in whether a cohort's social structure crystallizes early on after the beginning of college, or whether it takes longer to converge to a steady state.
We focus here on the ``cohort network,''
the set of all ties formed between individuals in the same college entry cohort, 
identified using the algorithm mentioned in the `Data` section. Since virtually all of its members
enter the institution at the same time, the cohort network is particularly meaningful as a unit of analysis
for network structure. Nonetheless, it is not the \textit{only} meaningful network one may investigate.
We may also talk of a ``co-presence'' network, composed of all individuals who are attending the college 
at the same time, as we could talk of the network composed of all individuals
who ever attended the university. The multi-cohort aspect of these networks makes their examination
more cumbersome, and we leave their investigation to future work. 

We investigate the 
convergence properties of a set of network statistics that reveal different aspects of the structure of each graph. For each 
class network we compute (1) the size of the largest connected component (relative to the total number
of students in the class), (2) the average ego-network clustering coefficient, (3) the modularity of the
modularity-maximizing partition obtained from the community detection algorithm presented by
Clauset, Newman and Moore \cite{clauset04}, and (4) the average shortest path length in the graph.
All network statistics were extracted using the SNAP network library \cite{leskovec2016snap}. The
statistics were extracted for every monthly time slice starting twelve months before the start date
we identified for the cohort, and ending five years after the start date. We expect months 0-45
to cover the typical time period associated with a 4-year undergraduate degree in the U.S. Statistics
obtained for the graph prior to this interval can yield insight into the networks preceding the college
experience proper, whereas the 15 months after will capture the period immediately following the
college experience for students who graduate within the typical four years. 

\begin{figure}
  \centering
  \includegraphics[width=\columnwidth]{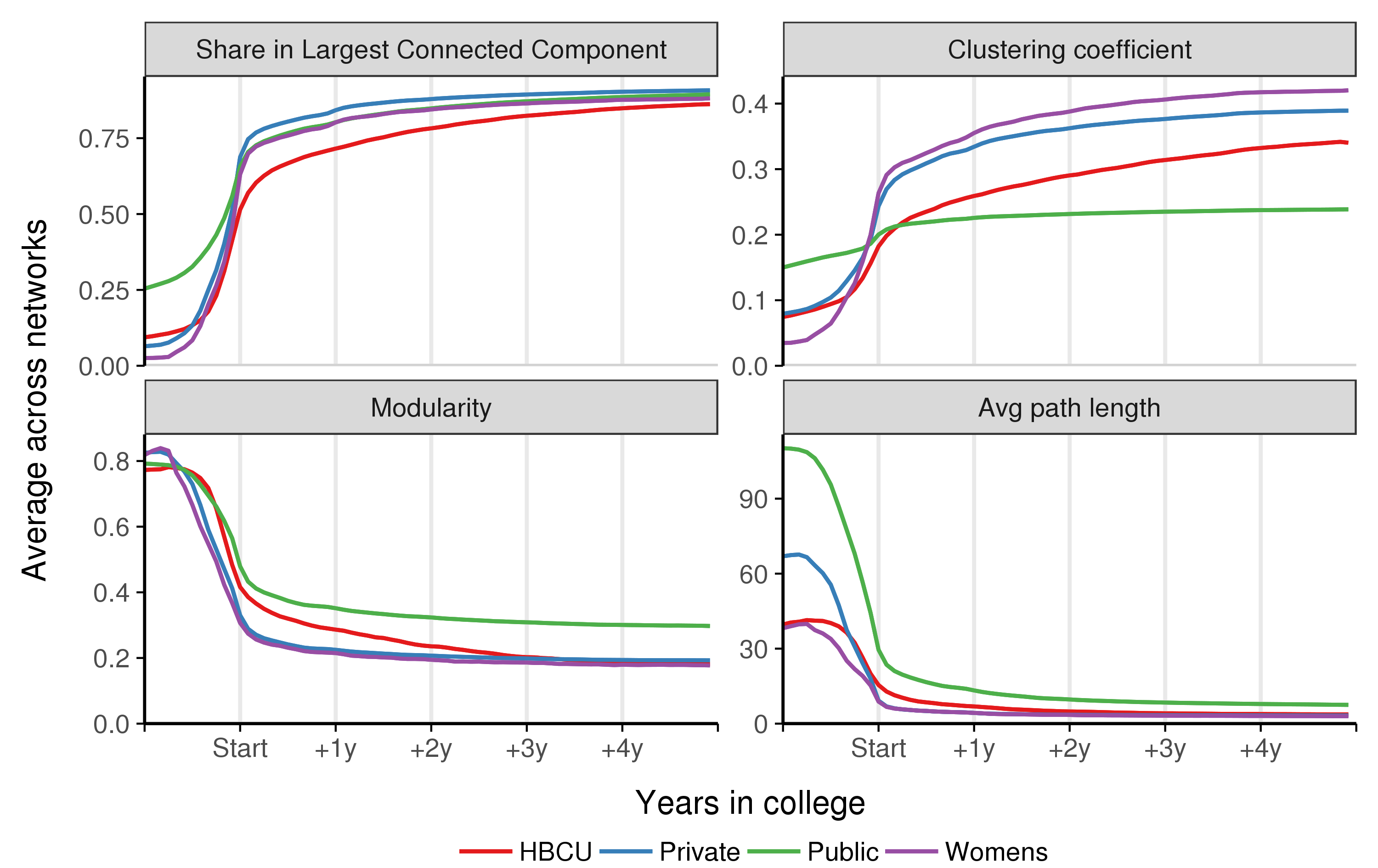}
  \caption{
    Average network statistics by number of months since the start of college by school type.
  }
  \label{fig:position_time}
\end{figure}

Monthly results for the aforementioned network statistics are averaged across all 7,586 entry year cohorts for each
of the 72 months in our observation window. Results are shown in Figure \ref{fig:position_time}, plotted separately
for private and public schools, as well as for HBCUs and for Women's colleges.\footnote{The categories mentioned here are exclusive, with minority-serving status taking precedence, i.e., a public HBCU would be considered an HBCU.}
The most immediately-striking feature of this figure is the extent to which the beginning 
of college changes the pre-college network existing between class members into something altogether different. The
moment school starts, we  see very sharp increases in the relative size of the largest connected component, and the
clustering coefficient, and sharp decreases in the modularity of the modularity-maximizing partition, and the average path
length connecting students. The network that exists prior to the beginning of college appears to be in a state where most
individuals are disconnected from one another (most of them not being in the largest connected component),
while those who are connected form a loose graph (with high average shortest path between members who are connected),
that has low clustering and is easy to partition (high modularity). Only a few months after, the graph changes dramatically:
the majority of students are in the largest connected component by the end of their first month, point at which we also
see the longest shortest path and modularity decrease markedly, while the clustering coefficient undergoes a significant
jump.

Another set of apparent differences from Figure \ref{fig:position_time} involves the public/private distinction. Public school
cohorts have a much larger connected component that pre-dates the start of university: on average, a quarter of students
at public schools are part of this connected component prior to school start, compared to only 6\% at private schools. Pre-existing
networks of public schools cohorts also show greater levels of clustering (15\%) compared to the homologous private school 
networks (8\%). By the first month of college the order is reversed. Public school cohorts now have more disconnected networks,
with 66\% of students being in the largest connected component, on average, compared to 68\% for private colleges and universities.
The same applies to the clustering coefficient, which at this point is 24\% for private schools, compared to 20\% for public schools.
After the first month, the clustering coefficient appears to continue increasing at a larger rate for private institutions than for public ones: 
by month 45 the average clustering coefficient is 38\% for private schools, and 24\% for public institutions of higher education.

Minority-serving institutions such as HBCUs and women's colleges also appear to display distinctive patterns. In particular,
HBCUs and women's colleges both show a stronger tendency towards increased clustering during the later college years.\footnote{This
sustained tendency to clustering is robust to examining public and private HBCUs separately. The analysis is not shown due to lack of space.} 
The size of the largest connected component appears smaller at HBCUs, which also display a more gradual decrease in modularity
over time.

\subsection{Stability of Node Positions}

In addition to examining overall network dynamics, the evolution of individuals nodes' position over time
provides another meaningful lens through which we can describe key trends in the life of a cohort social network.
We see centrality as a meaningful quantity not just from a graph theoretical perspective, but also as a measure of
the social capital accrued by an individual. In line with this expectation, related work on MBA students' social networks
has shown that the centrality of a student in their cohort's social network relates to professional success after
school~\cite{yang19}.

\begin{figure}
  \centering
  \includegraphics[width=\columnwidth]{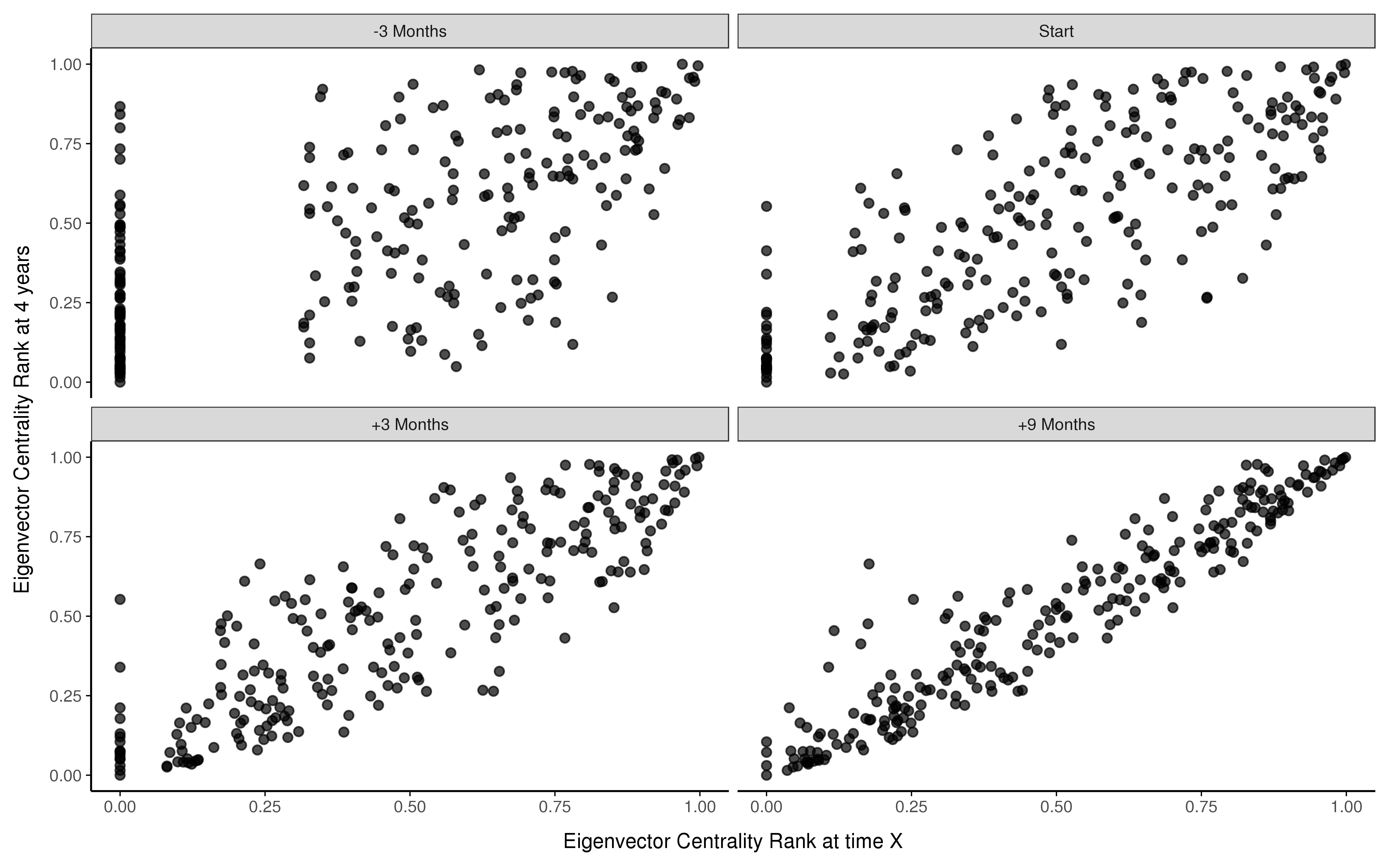}
  \caption{
    Eigenvector centrality rank at 45 months after the start of college, as compared to the same measure 
     at -3, 0, 3, and 9 months from the start of college, for 200 randomly-sampled individuals
     from a randomly-chosen cohort graph.
  }
  \label{fig:evcent_snapshots}
\end{figure}

With these considerations in mind, we are interested in understanding how quickly the centrality of nodes
in a cohort network stabilizes -- doing so is arguably essential for understanding the dynamic processes through
which social capital emerges as a resource during one's university years. We focus on eigenvector centrality, 
a measure that captures situations where differences in degree are meaningful measures of status~\cite{bonacich2007some},
which appears as a reasonable assumption in the social world of the university. Eigenvector centrality is computed
using the SNAP network library~\cite{leskovec2016snap} for every monthly snapshot of every cohort network. Resulting
network centrality scores are then rank-normalized within each monthly cohort network snapshot. The result is a node-level
score that is directly comparable between two snapshots of the same network. Figure~\ref{fig:evcent_snapshots} shows 
four such comparisons for a random sample of 200 nodes selected from a randomly-chosen cohort network. We compare
the rank-normalized centrality scores at months -3, 0, 3, and 9 with those at month 45, which we expect to be the
graduation date for most individuals in the cohort, given our focus on four-year undergraduate programs. We note 
the increase in the correlation coefficient from .53 between month -3 and month 45, to .7 when we consider month 0 against month 45,
to .82 for month 3, and .92 at month 9. We also note that after month 9, 
while some nodes do increase in centrality over time, there are hardly any nodes whose centrality markedly decreases.

\begin{figure}
  \centering
  \includegraphics[width=\columnwidth]{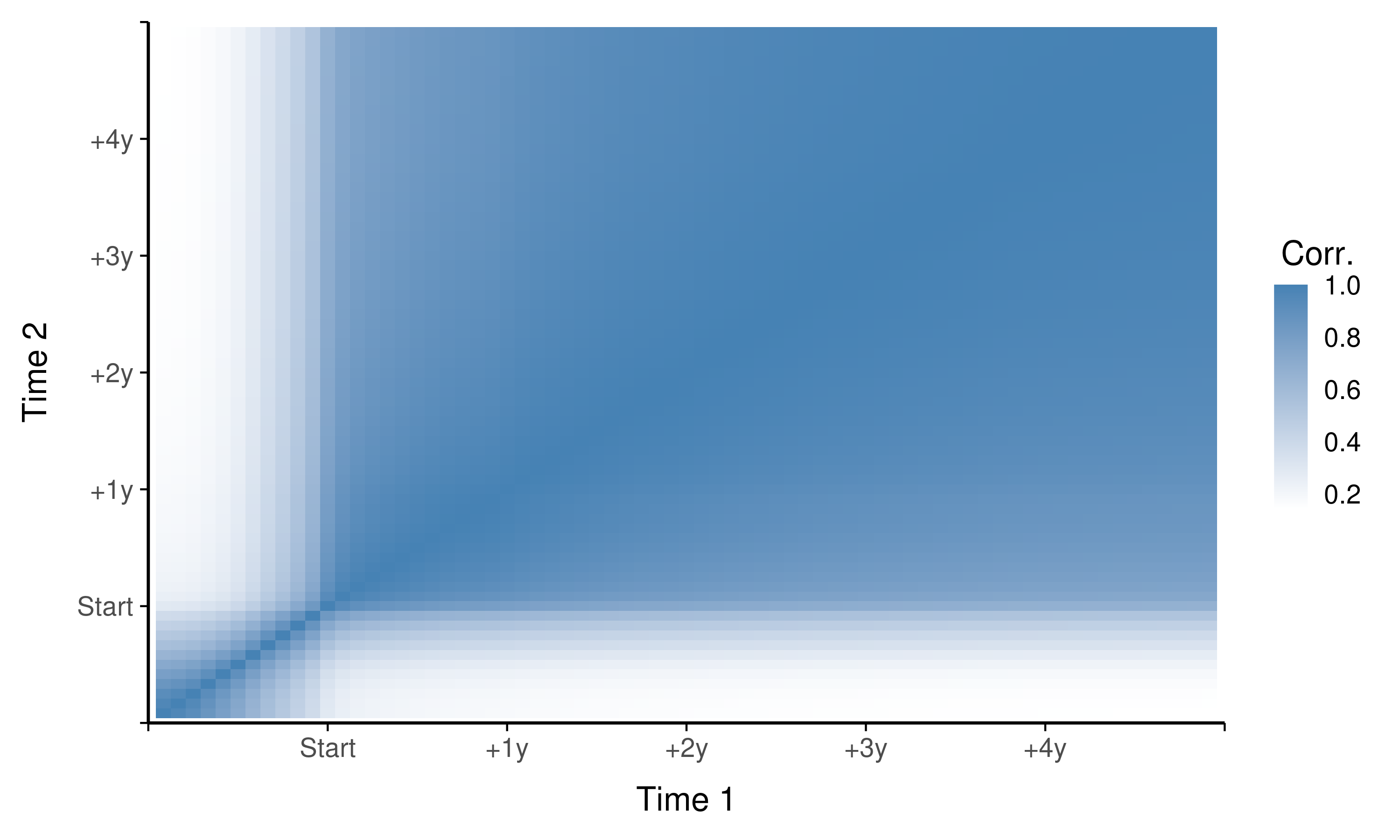}
  \caption{Correlation coefficients between ranked eigenvector centrality observed in every two months during the observation window.
    Correlations are computed for every cohort network, and then averaged across all cohort networks.}
  \label{fig:evcent_correlation_a}
\end{figure}

\begin{figure}
  \centering
  \includegraphics[width=\columnwidth]{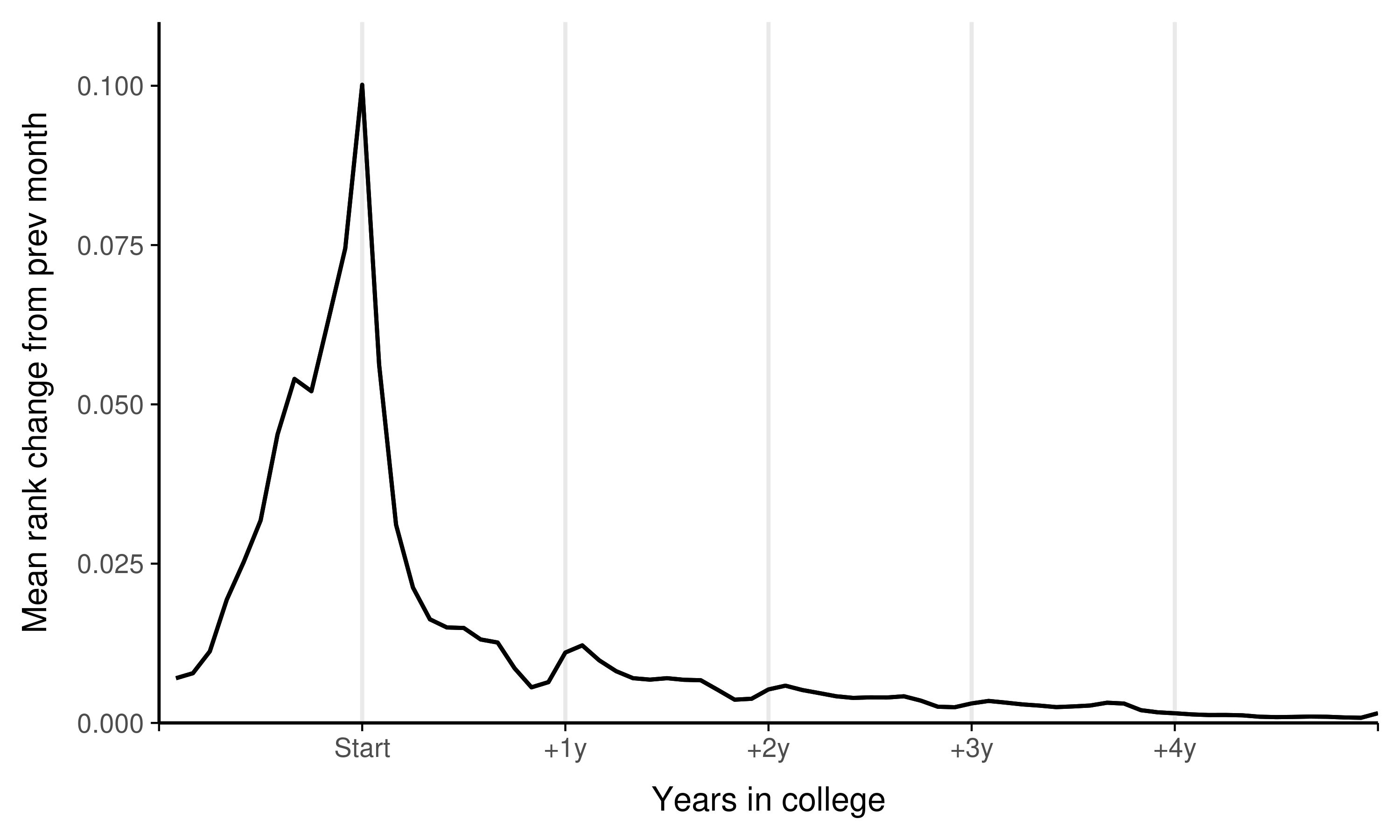}
  \caption{Average absolute month to month rank change within cohort networks of ranked eigenvector centrality. }
  \label{fig:evcent_correlation_b}
\end{figure}

We generalize the analysis of eigenvector centralities in Figure~\ref{fig:evcent_correlation_a}. To generate data for this
figure we start by computing correlation coefficients between rank-normalized eigenvector centrality scores for 
every pair of monthly snapshots of every cohort network. Correlation coefficients for a particular month pair
 are then averaged across all cohort networks, and the results are plotted as a heat map. The apparent block 
 structure of the heat map comes to suggest that for most cohort networks rank-normalized centrality during 
 the graduation month is highly correlated to centrality scores obtained during the first few months of college. This 
 result provides further evidence towards the inherent temporal stability of the structure of cohort social networks.

Despite the fact that the distribution of centralities is mostly stable after the first few months of college, shifts
in centrality \textit{do} occur for individual nodes. Figure \ref{fig:evcent_correlation_b} shows absolute average
month-to-month changes in rank in eigenvector centrality, averaged across all cohort networks. In addition to further
reinforcing the fact that most change happens during the beginning of the first year of college, the graph also
shows evidence that the amplitude of further shifts in centrality follows a seasonal pattern. While the absolute
rank change decreases from one year to the next, a ``bump'' in the amount of change is perceptible around
the start of each subsequent year, as well as at month 45, the likely graduation date for the cohort.

\section{Persistence of ties}

So far we have considered how ties form during a person's time in college. But which of these ties will remain active on Facebook
 after college and how does this relate to when and where they were formed? In this section we consider the present day closeness on Facebook
of two friends. We rank each person's Facebook friends by an aggregate measure modeled based on interaction frequency on the site, as computed in mid-2019. 
We then consider whether a college friend is in the top 200 friends for that person today, denoting such a tie as a 
``\textit{close(r) Facebook Friend},'' or CFF. Although measuring the strength of ties
via Facebook interactions is incomplete, it has been shown to correlate with survey questions about closeness~\cite{gilbert2009}.  

\begin{figure}
  \centering
  \includegraphics[width=\columnwidth]{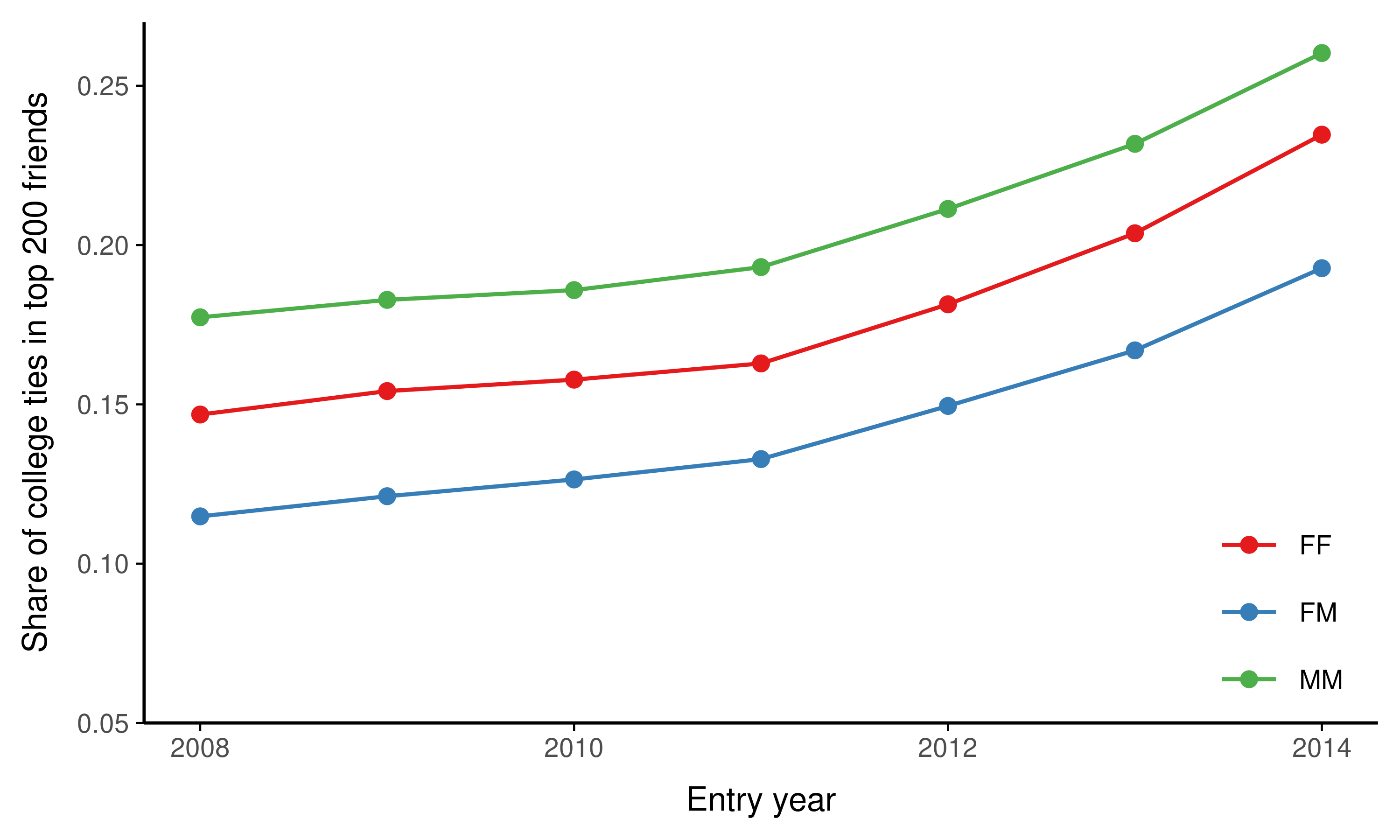}
  \vspace{-3mm}
  \caption{The share of college friendships that are in the top 200 Facebook friends (CFF), separated by starting year and gender. More recent ties and same-gender ties are more likely to be CFF.}
  \label{fig:persistence_time_gender}
\end{figure}

\begin{figure}
  \centering
  \includegraphics[width=\columnwidth]{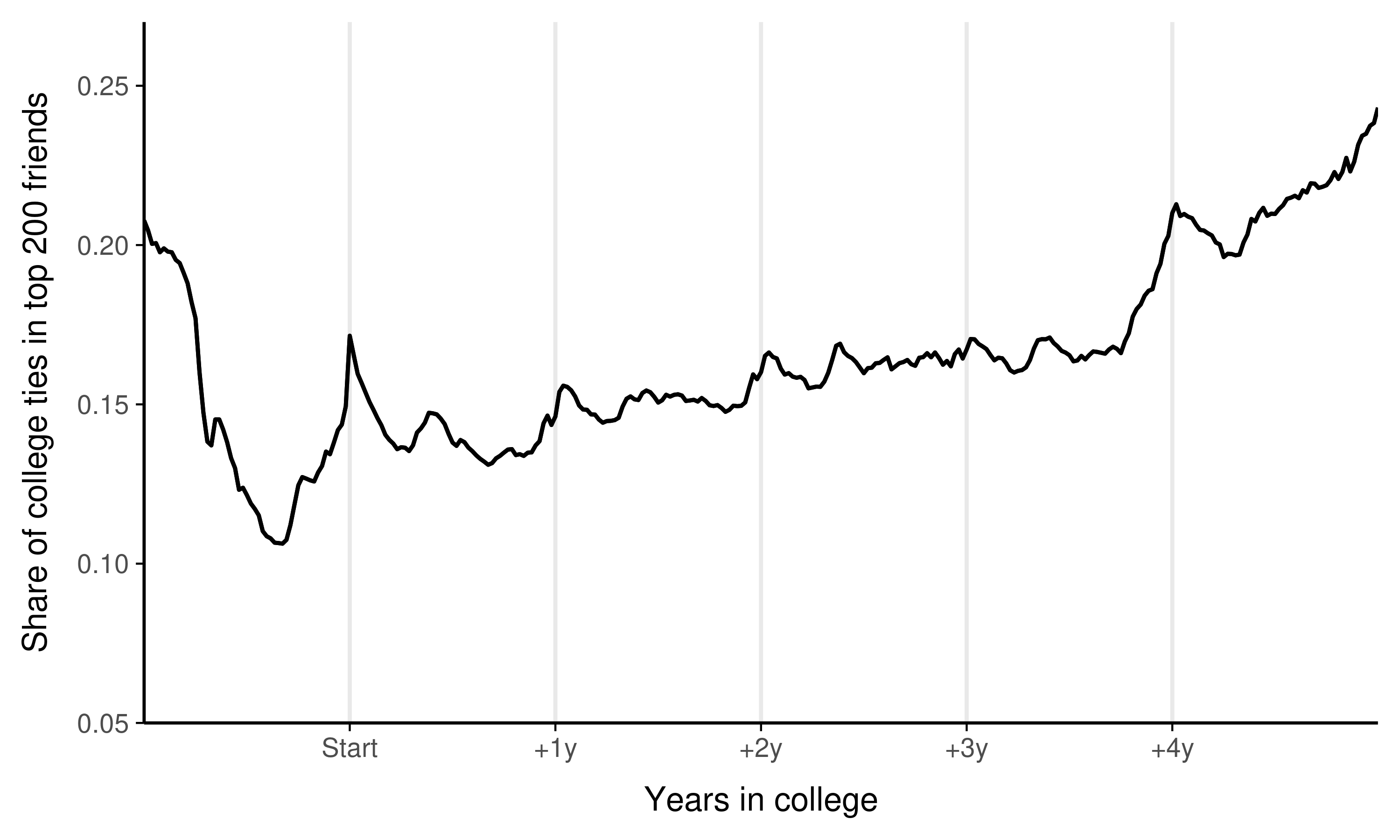}
  \caption{The share of college friendships that are in the top 200 Facebook friends (CFF), separated by week when the tie was formed relative to the start of school. These data exclude the last two entry-years, as they have a much stronger recency bias.}
  \label{fig:persistence_time}
\end{figure}

Figure~\ref{fig:persistence_time_gender} shows that time erodes the closeness of some college ties, as new ties are formed after college and not all ties continue interacting on Facebook. More recent cohorts have more of their college ties in their top 200, while successively older cohorts have fewer and fewer.
We further note differences by gender. On average, women have more Facebook friends than men, and they also have more college friends in their top 200 friends at present day. However, a lower \textit{share} of their college friends is CFF. Same-gender friendship are more likely to be CFF, with friendships between men being the most likely to stay within the top 200.

Earlier, we showed that formation patterns vary during different periods of the school year. Here we observe whether these different periods of tie formation correspond to differences in tie persistence.
Figure~\ref{fig:persistence_time} shows and overall mild upward trend in tie persistence that reflects the same effect we just mentioned: more recently formed friendships are more likely to still be active. Overlaid on this trend are several interesting features. Ties that pre-date college as well as admission decisions, are more likely to remain close than those formed close to but prior to enrollment. The oldest ties may be ones with high school friends that were reinforced when friends attended the same college. We also note a depression in share CFF in the months preceding the start of the first semester. These friendships may occur between people who form a tie because they both plan to enroll in the same college. However, once they enroll, they may not  have much additional shared context beyond attending the same college and so have lower tie strength than people who form the tie while in college. As seen in Figure~\ref{fig:new_edges}, this is also a time of more random edge formation, as measured by the share of edges that close triangles.

There is a substantial bump in tie strength at the very start of college followed by starts of subsequent years and also semesters.  One can only speculate why these ties are stronger. Potentially these are times that people meet new housemates with whom they will spend quite a bit of time, allowing for the formation of stronger ties. 
There is a marked jump in persistence for same-college ties formed \textit{after} college ends. This may be due to these ties having a persistent social context, such as graduate school or shared employment, after college.

\begin{figure}
  \centering
  \includegraphics[width=\columnwidth]{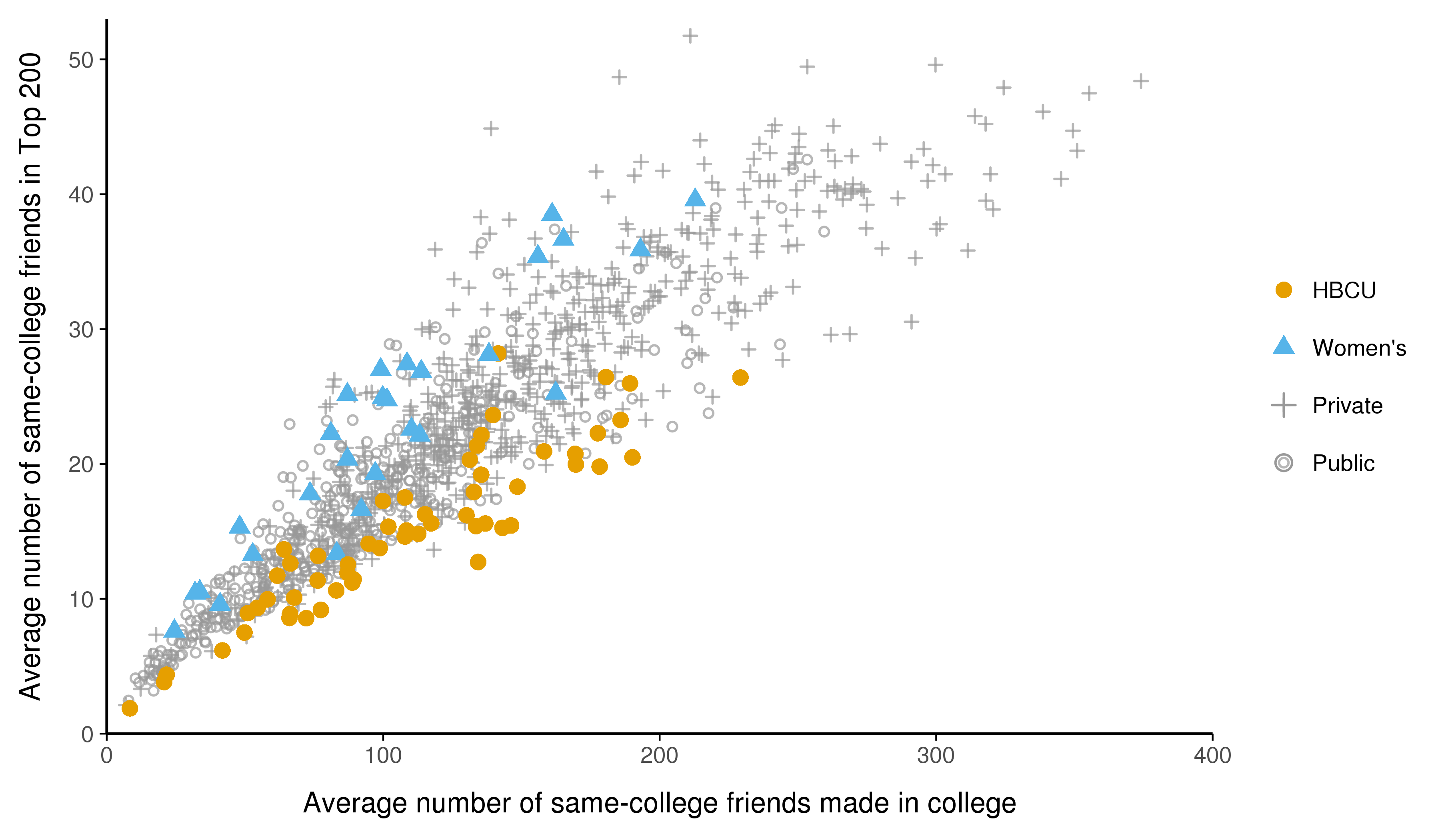}
  \caption{
      In colleges where people make many Facebook friends, the proportion of those ties which remain in the top 200 tends to be lower
      on average. Each data point represents the 2011 entering class at a college. Students at women's only institutions have a relatively larger share of CFFs and those at HBCUs a relatively smaller share.}
  \label{fig:persistence_sample}
\end{figure}

Earlier we described how one's network structure varies by the type of institution a person attends.
For example, people attending colleges where a majority of students live on-campus tend to friend more of their classmates on Facebook than those who attend commuter colleges. 
Similar heterogeneity can be observed in the long-term persistence of ties.
From the College Scorecard dataset, we group Historically Black colleges (HBCU), women's colleges, and then the remainder by whether they are private or public institutions.
As shown in Figure~\ref{fig:persistence_sample}, at institutions where people make more Facebook friends, and private colleges are more likely to fall in this group, people also tend to on average have a greater number of college friends among their closer Facebook ties ($\rho = 0.87, t= 152$).
Nevertheless, the likelihood of any particular tie being CFF is lower ($\rho = -0.38, t= 36$).
We note that some HBCUs fall slightly below the trend: 
given the number of Facebook ties formed in college, their proportion which remains CFF is lower.
For women's colleges the average trend is opposite: the ties are on average more likely to remain closer. This is consistent with Figure~\ref{fig:persistence_time}, showing that a higher proportion of same-gender ties are in the top 200. 

\begin{figure}
  \centering
  \includegraphics[width=\columnwidth]{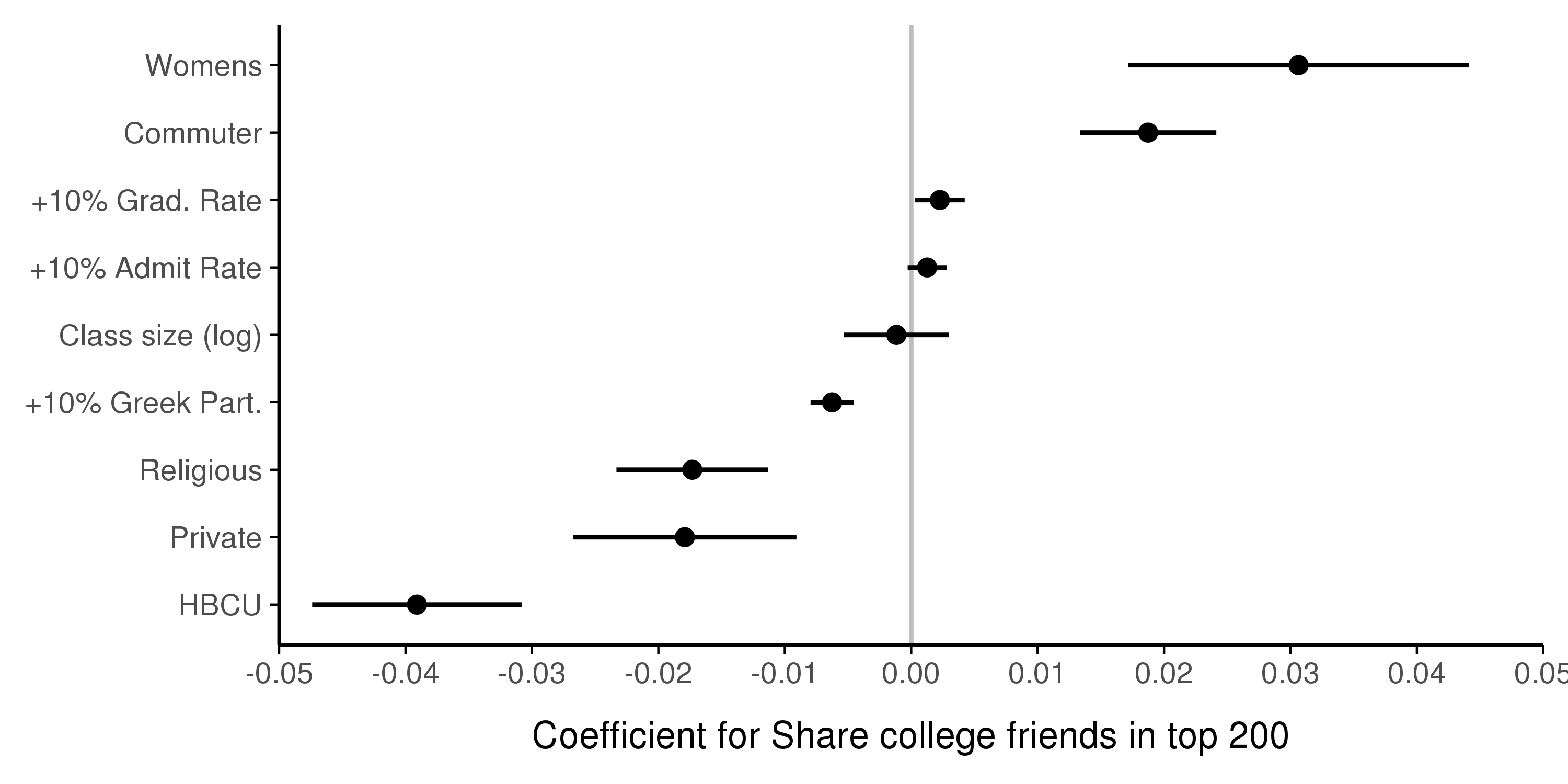}
  \caption{
    Estimates of differences in the share of college ties that are CFF by school covariates.
    Each data point represents an entry-class ($N=7,586$). Fixed-effects for year are not shown.
    Error bars correspond to 95\% confidence interval of the estimates. Standard errors are clustered by school.
  }
  \label{fig:persistence_coef}
\end{figure}

We systematize this analysis with a regression, with share CFF as the dependent variable, school characteristics as covariates, and fixed-effects for each year.
Each data point represents an entry-class ($N=7,586$).
The resulting estimates are shown in Figure~\ref{fig:persistence_coef} ($R^2=0.55$).
As discussed, women's-only schools form the highest share of CFF ties in college, about 3\% more when accounting for other covariates.
For commuter schools the story is a bit more complicated.
Students who attend dormitory schools make more Facebook friends at college, and a larger share of their CFFs are from college.
On the other hand, friendships started at commuter colleges are proportionally more likely to remain CFF.
One potential explanation is that residential colleges allow more context for many friendship ties to form, from sharing a residence or dining hall, to attending many on-campus social activities. However, these ties are also more likely to be incidental in nature.
Friendships formed at schools with more Greek activity are \textit{less} likely to stay CFF, as do those at religious schools and HBCUs.
Private schools also have about 2\% fewer proportional CFF's, as compared with public schools.
The effect here is similar to the commuter/dormitory story, as students at private schools have \textit{more} CFF's from college on average.

The above shows that when, where and how ties form has long term effects, at least for Facebook interactions. 
There are many factors which are not accounted for here, for example, the frequency with which people use Facebook, or the likelihood that they will continue their higher education, or change jobs or locations, all of which would potentially 
generate additional ties that would displace those made in college. We leave these and other questions for future work.

\section{Conclusion}

For a significant portion of the U.S. population,
  the social networks built during college are formative for the rest of life.
In this paper we sought to understand how college networks are shaped by different factors, such as homophily and propinquity, as well as institutional characteristics. The micro-level event data from Facebook permitted the study of these factors
  across a population of school networks covering most 4-year colleges in the U.S across a multi-year time-span.

We found evidence that the institutional context
  indeed mediates the formation of online social networks,
  and results into variety of network structure.
The influence of the educational ecology expresses itself in both time and college characteristics.
Edge formation spikes during times when students first have an opportunity to spend time together and become acquainted,
  for example at the start of a new academic period, which vary by school.
These are the moments when the networks change the most.
The demographic composition of one's social network is determined
  in part by the availability through the composition of the student body,
  and in part by homophily, the preference to connect with similar others.
This tendency is also affected by school characteristics.
Homophily by gender is generally lower in HBCUs,
  and varies in schools with high Greek participation,
  where same-gender edges are more prominent during Greek recruitment and lower otherwise.
Schools that attract students from afar spur the formation of ties between people from different hometowns,
  even in the period before the classes start.

On the other hand, the high-level structure of school networks does not undergo much
  change after the first months of school.
Incoming students immediately become part of the largest connected component, which along with the clustering coefficient, increases only modestly thereafter. Similarly, modularity and the average shortest path have the most marked drop at the start of enrollment.
One reason for this relative stability is the high density of these school graphs (the average density when school ends is 0.125),
  which leaves little room for the structure to change heavily after its initial formation.

We also looked at the persistence of ties after college, using an aggregated measure of tie closeness modeled based on activity on the site.
Intuitively, more recent ties are more likely to still be close.
However, ties formed well before college starts are more likely to stay close than those formed during the period immediately preceding the start of school.
Tie persistence also varies by school characteristics, with ties formed in residential and women's-only schools more likely to stay close.
When and where ties form thus has long term effects on tie persistence on Facebook.

\subsection{Limitations and future work}

While this paper presents an exploratory look into the formation of social networks in college, it does not fully model the factors shaping the formation of ties.
Though we identified some of the ecological factors that affect network formation,
  there are many other aspects of the educational experience that have been shown to be instrumental,
  including what classes students took \cite{kossinets06},
    their extra-curriculars \cite{vanduijn03,schaefer11},
    and, where relevant, their dormitory \citep{festinger50,marmaros06,lewis12,vanduijn03},
  which are not studied in this paper.
The same holds for aspects of student demographics,
  like ethnicity \cite{currarini10,wimmer10,joyner00} and socioeconomic status \cite{goldrickrab06}.

Our analysis is likely affected by the need to approximate, for those individuals who
  did not specify their years in college, when they started and stopped attending.
Both the accuracy of our year assignment procedure and the four-year graduation rate,
  are correlated with college characteristics,
  which affects how the data is constructed.
Our analysis applies to social networks on Facebook only.
While other work has argued that Facebook networks mirror those offline \cite{gilbert2009},
  at least structurally \citep{dunbar15,arnaboldi12,bailey18},
  others have observed that the amount of activity on Facebook is correlated with the  number of Facebook friends \citep{lewis08,viswanath09}.
Our findings are possibly confounded by changes in the design of the Facebook platform, but the results are mostly stable across the different starting cohorts.

All analysis in this work is observational.
Future studies could attempt to leverage natural experiments \cite{phan15}, or randomized assignment \cite{carrell13} to identify causal effects.
We leave the mechanisms behind any correlations, and the extent to which they are confounded by other factors, like Facebook activity, for future work.

What further factors drive the formation of particular edges is another rich dimension of college networks
  that still has to be explored using one of the many available methods \cite{snijders01,leskovec08,overgoor19a}.
Finally, social networks have been shown to affect individuals' outcomes, such as employment, throughout life. One could study the properties of ties originating in college after college ends and
 potentially compare structural measures to known measures of social mobility,
 like the ones prepared by \cite{chetty17}. We leave these and other questions to future work.

\begin{acks}
The authors thank Johan Ugander and Eduardo Laguna-M{\"u}ggenburg,
  for helpful comments and feedback.
\end{acks}

\bibliographystyle{ACM-Reference-Format}
\bibliography{scorecard_time}

\end{document}